\documentclass[3p, twocolumn]{elsarticle}

\usepackage{lineno,hyperref,color,multirow,paralist,stfloats, graphicx}
\usepackage{makecell}
\usepackage{bbding}
\modulolinenumbers[5]

\journal{Journal of Ad Hoc Networks}

\begin{document}

\begin{frontmatter}

\title{Beyond Beaconing: Emerging Applications and Challenges of BLE}


\author[aut_1_addr]{Jian Yang\corref{correspondingauthor}}
\cortext[correspondingauthor]{Corresponding author}
\ead{jyang9@nd.edu}

\author[aut_1_addr]{Christian Poellabauer}
\ead{cpoellab@nd.edu}

\author[aut_2_addr]{Pramita Mitra}
\ead{pmitra3@ford.com}


\author[aut_2_addr]{Cynthia Neubecker}
\ead{cneubeck@ford.com}

\address[aut_1_addr]{Department of Computer Science and Engineering, University of Notre Dame, Notre Dame, IN 46556, USA}
\address[aut_2_addr]{Research and Advanced Engineering, Ford Motor Company, Dearborn, MI 48121, USA}

\begin{abstract}
As an emerging technology with exceptional low energy consumption and low-latency data transmissions, Bluetooth Low Energy~(BLE) has gained significant momentum in various application domains, such as Indoor Positioning, Home Automation, and Wireless Personal Area Network~(WPAN) communications. With various novel protocol stack features, BLE is finding use on resource-constrained sensor nodes as well as more powerful gateway devices. Particularly proximity detection using BLE beacons has been a popular usage scenario ever since the release of Bluetooth 4.0, primarily due to the beacons' energy efficiency and ease of deployment. However, with the rapid rise of the Internet of Things~(IoT), BLE is likely to be a significant component in many other applications with widely varying performance and Quality-of-Service (QoS) requirements and there is a need for a consolidated view of the role that BLE will play in applications beyond beaconing. This paper comprehensively surveys state-of-the-art applications built with BLE, obstacles to adoption of BLE in new application areas, and current solutions from academia and industry that further expand the capabilities of BLE.
\end{abstract}

\begin{keyword}
Bluetooth Low Energy \sep BLE \sep Communication\sep Applications\sep Low Power\sep Low Latency.
\end{keyword}

\end{frontmatter}

\section{Introduction}
Bluetooth Low Energy~(BLE), also known as Bluetooth Smart, is an emerging short-range wireless technology aiming at low-power, low-latency, and low-complexity communications. With its deep market penetration (e.g., Bluetooth is available on almost all laptops, tablets, and smartphones), BLE has become an attractive alternative to many existing wireless communications technologies~\cite{tosi2017performance}. A new implementation of the Bluetooth protocol stack allows BLE to operate for very long time periods using only a coin-cell battery~\cite{collotta2015bluetooth}. BLE also provides new approaches to wireless communications compared to the Bluetooth Basic Rate/Enhanced Data Rate (BR/EDR) supported by Classic Bluetooth, as described later in the paper.

One of the most popular BLE applications is its use in {\em BLE Beacons}, which is a class of devices that continuously broadcast an identifier to nearby BLE receivers~\cite{ gast2014building}. This has enabled a multitude of {\em proximity detection} solutions, i.e., the beacons allow devices such as smartphones and tablets to perform certain actions when in close proximity to a beacon. Examples of using BLE Beacons include indoor positioning~\cite{lin2015mobile,faragher2015location}, activity recognition~\cite{de2015multimodal, vzidek2013smart}, and vehicle network wake-up system~\cite{takahashi2015evaluation} .

\begin{figure*}[htbp]
\centering
\includegraphics[width=5.2in]{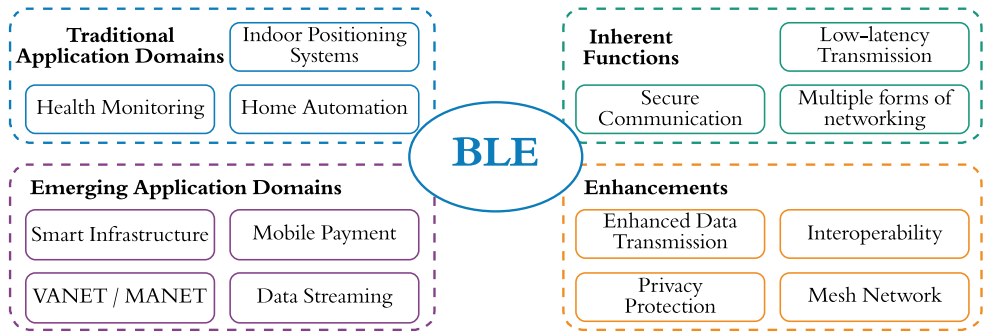}
\caption{Taxonomy of BLE applications and their challenges and solutions. }
\label{fig_taxo}
\end{figure*}

However, in addition to proximity detection, BLE-based systems are also used to connect wireless sensors and receivers in a variety of healthcare and smart home applications. Because of its ubiquitous presence in commercial devices and low-energy requirements, recently a few less traditional domain of applications are investigating the use of BLE, such as Vehicular Ad Hoc Networks~(VANETs)~\cite{bronzi2016bluetooth,yang2017using}, smart infrastructure~\cite{kim2016navigating,rajagopal2014low,vochin2017intelligent}, multimedia streaming devices~\cite{giovanelli2015bluetooth,meli2015streaming,shao2016years}, mobile payment systems~\cite{todasco2015systems,baldie2016system}, etc. However, such emerging applications are often unable to utilize BLE without modification or enhancements. Obstacles to the widespread adoption of BLE (besides beaconing) include the lack of support for large and dynamic data transmissions, mesh networking, inter-operability with other wireless technologies, and security/privacy protection. 

While there have been solutions proposed, both in academia and industry, to address these challenges, there are still many remaining opportunities to ensure BLE's success in different domains. This paper provides a comprehensive review of emerging application domains of BLE, the challenges BLE faces in these domains, and solutions that have been proposed. Figure~\ref{fig_taxo} presents a taxonomy of BLE applications and their supporting techniques that will be discussed in this paper. The inherent functions of BLE provide the basis for both {\em primary} and {\em emerging} applications, while emerging applications often require enhanced features that have not yet been developed. Therefore, we will first introduce the inherent functionality of BLE, including the {\em low-latency} capability, {\em security} consideration for communication, and basic {\em network topology} that BLE supports. For each of the application domain, we review several typical examples. Then we focus on recent proposed BLE enhancements regarding challenges in the areas of {\em data transmission}, {\em mesh networks}, {\em interoperability}, and {\em privacy protection}.

The rest of this paper is structured as follows. In Section~\ref{overview}, we introduce the primary functions of BLE and highlight its most important features provided by various protocol revisions. Section~\ref{app} summarizes and discusses emerging application domains and their characteristics. Then in Section~\ref{sol}, we present a comprehensive survey of existing modifications and enhancements of BLE for specific application scenarios and discuss open challenges and issues for future BLE applications. Finally, we conclude the paper with a review of our insights in Section~\ref{con}.

\section{BLE: Current Functions and Revisions}\label{overview}

BLE was first introduced in 2010 by the Bluetooth Special Interest Group~(SIG) as part of the Bluetooth 4.0 specification~\cite{ble4_0}, which defined the overall architecture and implementation details of BLE. Since then, there have been several revisions of the Bluetooth core specification: Bluetooth 4.1~\cite{ble4_1}, Bluetooth 4.2~\cite{ble4_2}, Bluetooth 5.0~\cite{ble5}, and Bluetooth 5.1~\cite{ble5_1}. Major improvements by these revisions address power management, throughput, communication range, latency, and security issues with BLE. This section describes the primary architecture, current functions, and major evolution of BLE.

\subsection{BLE Stack}

\begin{figure}[htbp]
\centering
\includegraphics[width=3in]{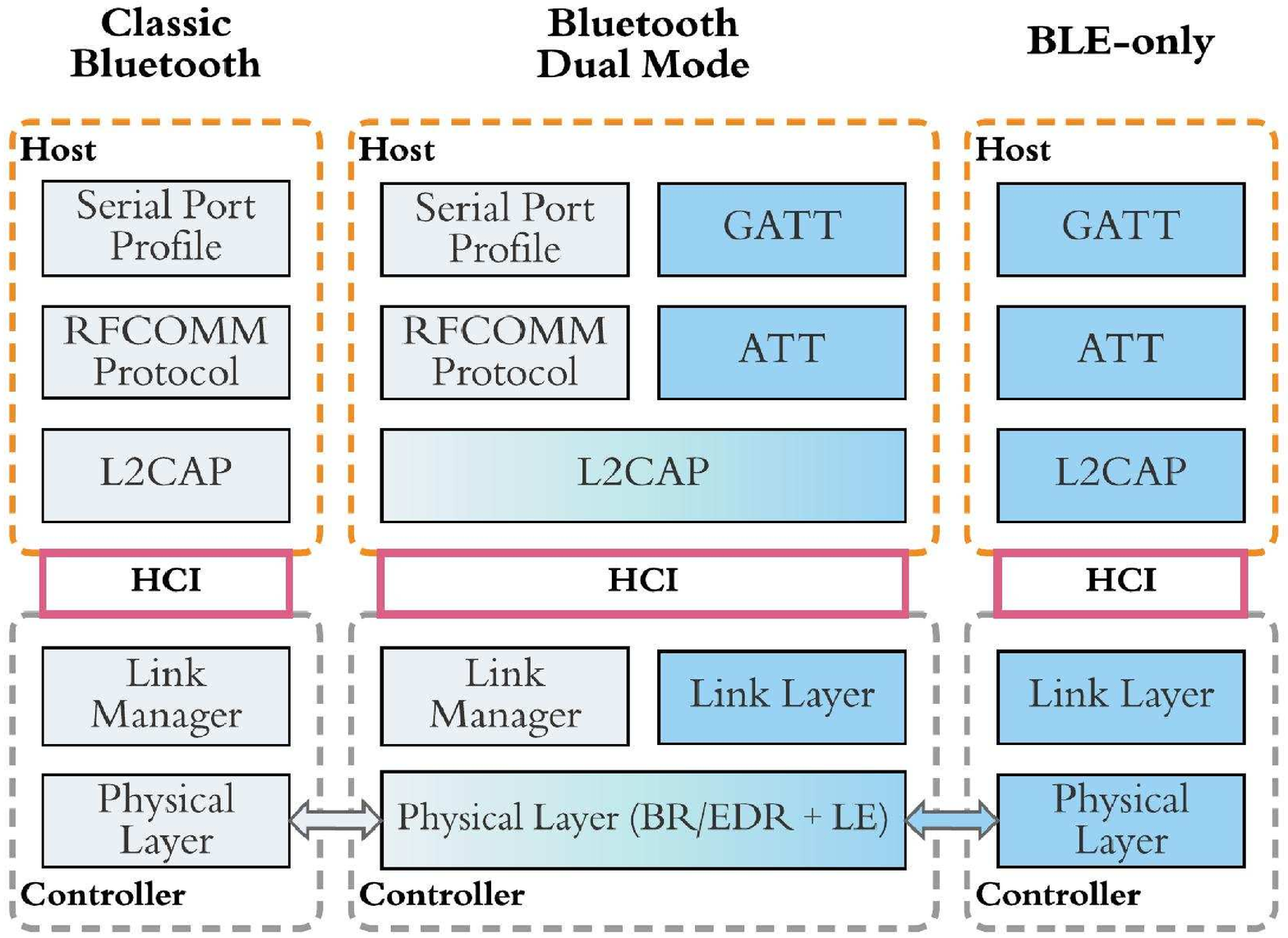}
\caption{A simplified protocol stack for three types of Bluetooth chipsets (adapted from~\cite{galeev2011bluetooth}).}
\label{fig_ble_stack}
\end{figure}

The protocol stack of BLE maintains a similar lower layer structure as Classic Bluetooth, but also provides revised implementations and a few new layers, such as the {\em General Attribute Profile~(GATT)} and the {\em Attribute Protocol~(ATT)}. Figure~\ref{fig_ble_stack} shows the simplified stack for three major types of Bluetooth chipsets: Classic Bluetooth, Dual Mode, and BLE-only. For compatibility reasons, the {\em Bluetooth Dual Mode} chipsets were introduced to support both Low Energy and BR/EDR communications, which are commonly seen on phones and tablets. BLE-only chipsets are typically installed on cost- and resource-constrained devices.

The implementation of the BLE stack layers focuses on low-latency and low energy consumption. Several highlights of the stack layers can be summarized as follows. First, the {\em Physical (PHY) Layer} of BLE defines 40 Radio Frequency (RF) channels in the 2.4 GHz band, among which three channels are defined as advertising channels for disseminating data and 37 data channels are used for bidirectional exchange of messages in established connections.  
The {\em Link Layer (LL)} defines two interaction patterns between two devices: 1) {\em connection-less} communication, i.e., two devices act as {\em advertiser} and {\em scanner}, where the advertiser broadcasts data packets and the scanner can receive them; 2) {\em connection-based} communication, i.e., the scanner and the advertiser are able to establish a bidirectional connection and adopt the role of {\em central} and {\em peripheral}, respectively. The Host Controller Interface~(HCI) is a standard protocol that takes care of the communication between the {\em Host} and {\em Controller}.

On top of the HCI sits the {\em Logical Link Control and Adaptation Protocol~(L2CAP)}, which is a protocol in common with Classic Bluetooth. It handles the data from lower layers and encapsulates them into the standard BLE packet format for upper layers' use and vice versa.
The serial port protocols ({\em Serial Port Profile} and {\em RF Communication Protocol}) in Classic Bluetooth are replaced by two novel layers in BLE stack: the {\em ATT} and {\em GATT}. These two layers handle the data interaction between application and stack layers when connection-based BLE communications are used~(Section~\ref{data_trans}). Furthermore, the {\em Generic Access Profile~(GAP)} and {\em Security Manager Protocol~(SMP)} (not shown) handle the general link management and security functions in BLE communications.

\subsection{Built-in Functions}
BLE was originally designed for applications in the rising IoT industry where Classic Bluetooth may be found less efficient. The new implementation of the BLE stack protocol inherently provides novel features with respect to data transmission, networking, and security. In this subsection, we discuss these built-in features of BLE and how they can be used to support various application domains.

\subsubsection{Data Transmission}\label{data_trans}

\begin{figure}[htbp]
\centering
\includegraphics[width=3in]{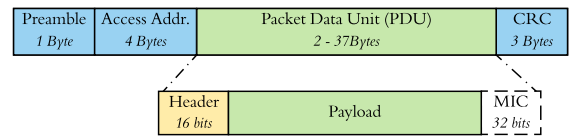}
\caption{The data packet format.}
\label{fig_data_packet}
\end{figure}

The Link Layer defines one packet format (Figure~\ref{fig_data_packet}) used for both advertising packets (also known as {\em advertisements}) and data channel packets, which support connectionless and connection-based communication, respectively. These two types of communication dispense with the slow and low-responsive pairing mechanism used in Classic Bluetooth, which makes the protocol more attractive to several novel application domains. 

The {\em Preamble}, {\em Access Address}, and {\em Cyclic Redundancy Check~(CRC)} fields are defined for easy control by the Link Layer, with regard to different communication scenarios. Data is transmitted via the {\em Packet Data Unit~(PDU)} payload, with the {\em Header} identifying its type and size.

In connectionless communications, data is broadcast directly via advertisement payloads and all neighbors (i.e., nodes in wireless range) can receive it. Since the payload is limited to a certain size (31 bytes in Bluetooth 4.1), a single advertisement can transmit only very short messages. Splitting and reassembling data for multiple advertisements can help with large data transmissions, but there will be a trade-off between data volume, transmission overhead, and packet loss. Therefore, connectionless communication is typically used for BLE beaconing, where beacons operate as advertisers and continuously broadcast a specific packet containing information for various purposes. The advertising process consumes very little energy and the theoretical life of a BLE beacon powered by a coin cell battery can be up to 14 years~\cite{gomez2012overview}. 

An advertiser is able to announce its availability for connections by setting the PDU header in outgoing advertisements accordingly, so that the receiving scanner can initiate a connection request. 
A connection will be established once the advertiser receives and accepts the connection request, and then the advertiser becomes a peripheral and the scanner becomes a central. The data exchange will then be processed based on the GATT rules via one of the 37 data channels. 

Connection-based communication uses the same packet format, with the {\em Message Integrity Check~(MIC)} field used for encrypted transmission checks. The {\em adaptive frequency hopping mechanism} is applied to data channels, with the purpose of minimizing channel interference and maintaining a low loss rate. The GATT further defines the rules of exchanging data. The peripheral acts as a {\em GATT server} that stores data in the form of a set of services. Each service contains a number of characteristics that take the length of a single PDU. The connected central device, known as {\em GATT client}, is able to discover, read, and write to the services and characteristics according to their permissions. As an example, if we have a heart rate monitor acting as a peripheral, its GATT server will contain a \textit{Heart Rate Sensor} service with a \textit{Heart Rate} characteristic. This characteristic will include the sensor data and the access permissions (e.g., [60bpm, Read Only]).

\subsubsection{Networking}
The central-peripheral relationship of connection-based BLE communications is similar to the master-slave relationship in BR/EDR communications, where a central device is allowed to connect to multiple peripherals. Therefore, this communication mode supports the most basic network topology, a {\em piconet} that is composed of one central and multiple peripherals. Unlike BR/EDR slaves, BLE peripherals do not share a common physical channel with the central, i.e., each peripheral communicates on a separate physical channel with the central. Since Bluetooth 4.1, each device has the capability to operate simultaneously in both roles in different piconets, thus providing opportunities to form larger {\em scatternets}. 

Furthermore, the connectionless communication operates on different physical channels (three advertising channels), and can coexist with connection-based communications (on data channels). On one hand, advertisers and scanners can form a broadcast network on advertising channels; on the other hand, the central or peripheral in an existing piconet or scatternet are also able to advertise and scan, resulting in a scatternet that involves both advertising and data channels. There could be multiple combinations of such scatternets.

A more complex network topology, which offers path diversity that can cope with radio propagation impairments and node failures, is the mesh network. This type of network has been introduced in BLE specifications starting from version 4.2, where a node can act as central and peripheral simultaneously. In late 2017, the Bluetooth SIG released a set of specifications and profiles that define BLE mesh networking on top of current BLE stacks~\cite{bluetooth2017mesh}. However, no implementations or tests of mesh networks have yet been provided by the group. Related problems, such as dynamic address allocation, network topology mapping, and routing, also require further exploration.

\subsubsection{Security}
In general, Bluetooth has potential vulnerabilities that fall into three categories: 1) passive eavesdropping, 2) man in the middle (MITM) attacks, and 3) identity tracking.

Passive eavesdropping is the process where a third device listens to the data being exchanged between two connected devices. BLE addresses this by encrypting the data being transferred using AES-CCM encryption at the Link Layer. The AES-CCM encryption is considered secure as long as the key is unpredictable.  

MITM attacks occur when a malicious device impersonates two other legitimate devices, in order to fool these devices into connecting to it. In this scenario, both the central and peripheral will connect to the malicious device, which in turn routes the traffic between the two other devices. This gives the legitimate devices the illusion that they are directly connected to each other when in fact their connection has been compromised. This setup not only allows the malicious device to intercept all the data being sent, but also allows it to inject false data into the communication or remove data before it reaches its intended recipient. LE Secure connections were introduced in version 4.2 to address this problem, where three steps are required before exchanging data: 1) a pairing feature exchange, 2) key generation via Elliptic Curve Diffie Hellman (ECDH) encryption, and 3) authentication.

Finally, identity tracking is where a malicious entity is able to associate the address of a BLE device with a specific user and then physically track that user based upon the presence of the BLE device. BLE addresses this by periodically changing the device address to make it untraceable.

\subsection{BLE Versions}
Since the release of version 4.0, several revisions of the Bluetooth specification have been published, with enhancements in data rate, payload, power consumption, and security. Table~\ref{ble_comp} summarizes the major differences between these versions.

\begin{table}[thbp]
\caption{Comparison of Bluetooth versions.}
\label{ble_comp}
\resizebox{\columnwidth}{!}{%
\begin{tabular}{l|c|c|c|c}
\hline
\textbf{Version} & \textbf{4.0} & \textbf{4.1} & \textbf{4.2} & \textbf{5.0 \& 5.1} \\ \hline
\textbf{Multi-Roles} & No & \multicolumn{3}{c}{Yes} \\ \hline
\textbf{\begin{tabular}[c]{@{}l@{}}PDU\\ Payload\end{tabular}} & \multicolumn{2}{c|}{\begin{tabular}[c]{@{}c@{}}Up to \\ 31 bytes\end{tabular}} & \multicolumn{2}{c}{\begin{tabular}[c]{@{}c@{}}Up to\\ 255 bytes\end{tabular}} \\ \hline
\textbf{LE Secure} & \multicolumn{2}{c|}{No} & \multicolumn{2}{c}{Yes} \\ \hline
\textbf{\begin{tabular}[c]{@{}l@{}}IoT\\ Support\end{tabular}} & \multicolumn{2}{c|}{Limited} & Medium & High \\ \hline
\textbf{\begin{tabular}[c]{@{}l@{}}Advertising\\ Channels\end{tabular}} & \multicolumn{3}{c|}{3 Channels} & \begin{tabular}[c]{@{}c@{}}3 Primary Ch.\\ 37 Secondary Ch.\end{tabular} \\ \hline
\textbf{Data Rate} & \multicolumn{3}{c|}{1 Mbps} & 2 Mbps \\ \hline
\textbf{\begin{tabular}[c]{@{}l@{}}Effective\\ Range\end{tabular}} & \multicolumn{3}{c|}{\begin{tabular}[c]{@{}c@{}}50 m (Line of Sight)\\ 10 m (Indoor)\end{tabular}} & \begin{tabular}[c]{@{}c@{}}200 m (Line of Sight)\\ 40 m (Indoor)\end{tabular} \\ \hline
\textbf{Battery Life} & \multicolumn{3}{c|}{Shorter} & Longer \\ \hline
\end{tabular}%
}
\end{table}

Bluetooth 4.0 explicitly prohibits a peripheral to participate in multiple connections (or assume multiple roles) simultaneously with other central devices. Version 4.1 and later incorporate a fundamental change with regard to the roles each device can play when multiple connections are present. That is, a device, regardless of its Link Layer role, can run multiple Link Layer instances simultaneously without limitation. Therefore, a peripheral is allowed to be simultaneously connected to more than one central device. 

Version 4.2 introduces major enhancements where the maximum PDU payload can be up to 255 bytes, compared to only 31 bytes in previous versions.
This version also provides additional support for IoT capabilities, such as low-power IP~\cite{yim2015ipv6} and Internet gateways~\cite{decuir2014introducing}. In terms of security, the \textit{LE Secure Connections} were introduced in this version, which utilize the ECDH algorithm for key generation, in order to protect against MITM attacks. 

Bluetooth 5.0 is regarded as a significant leap forward compared to previous versions, with claims such as ``twice the speed'' and ``four times the range''~\cite{ble5}. This version also defines two types of advertising channels: {\em primary} and {\em secondary}. The primary advertising channels are the same three advertising channels available in previous versions, while the secondary advertising channels use the remaining 37 BLE channels (formerly defined solely as data channels). The secondary advertising channels can exploit frequency hopping, just like data channels. The recently released Bluetooth version 5.1 provides further improvements in connection latency and location services to better support indoor localization services. Both versions 5.0 and 5.1 were designed for IoT applications, e.g., by using more advanced power management designs to maximize the life time of battery-powered devices.

\section{Application Domains}\label{app}

BLE's low-energy performance and widespread deployment in mobile devices makes it an excellent candidate for a variety of applications, including many emerging application domains. Traditional application scenarios can take advantage of all inherent BLE functionality immediately and include examples found in Indoor Positioning Systems~(IPS), health monitoring, and home automation. In contrast, many emerging applications cannot readily use BLE and thereby may require further modifications or enhancements, including applications in smart infrastructure, vehicular networks, mobile payments and data streaming. While BLE beacons only represent one specific realization of the BLE technology (and previous work has investigated its use in IoT applications in depth~\cite{jeon2018ble,posdorfer2016towards}), this paper focuses on other forms of BLE technology, such as Bluetooth Dual Mode or BLE-only chipset. In this section, we briefly summarize the traditional application domains of BLE, and review some attempts of using BLE in emerging domains. 

\subsection{Traditional Application Domains}
BLE was designed specifically for static applications with low-energy requirements.
In order to support quick and simple development, each BLE chipset comes with a built-in list of uniform type identifiers, which cover some basic services or data types for IPS, health monitoring, and home automation. These are the three main application domains where BLE has increasingly been deployed and evaluated. Table~\ref{tab:trad_app} summarizes the built-in BLE features used in these domains.

\begin{table}[htbp]
\centering
\caption{Summary of Traditional BLE Application Domains.}
\label{tab:trad_app}
\resizebox{\columnwidth}{!}{%
\begin{tabular}{c|c|c|c}
\hline
\textbf{\begin{tabular}[c]{@{}c@{}}App \\ Domain\end{tabular}} & \textbf{\begin{tabular}[c]{@{}c@{}}BLE \\ Built-in Features\end{tabular}} & \textbf{\begin{tabular}[c]{@{}c@{}}Range of \\ Interest\end{tabular}} & \textbf{Examples} \\ \hline
IPS & \begin{tabular}[c]{@{}c@{}}connectionless communication, \\ low energy\end{tabular} & \begin{tabular}[c]{@{}c@{}}$< 30 m$\\ (workspace)\end{tabular} & \cite{faragher2015location, vcabarkapa2015comparative, altini2010bluetooth, palumbo2015stigmergic, kriz2016improving} \\ \hline
\begin{tabular}[c]{@{}c@{}}Health \\ Monitoring\end{tabular} & \begin{tabular}[c]{@{}c@{}}GATT profile, \\ piconet networking, \\ privacy protection\end{tabular} & \begin{tabular}[c]{@{}c@{}}$< 1.5 m$ \\ (body area)\end{tabular} & \cite{ali2011bluetooth, guo2013design, lin2014bluetooth, zhou2013bluetooth} \\ \hline
\begin{tabular}[c]{@{}c@{}}Home \\ Automation\end{tabular} & \begin{tabular}[c]{@{}c@{}}GATT profile,\\ piconet networking, \\ privacy protection\end{tabular} & \begin{tabular}[c]{@{}c@{}}$< 15 m$\\ (living area)\end{tabular} & \cite{khan2016context, collotta2015bluetooth, collotta2015novel, galinina2015smart} \\ \hline
\end{tabular}%
}
\end{table}

IPS are systems that can determine the position of an object or a person in a limited physical space~\cite{mainetti2014survey}. A simplified structure of an IPS is based on the periodical execution of two steps: 1) sensors or receivers receive signals from transmitting devices and 2) distributed devices or a central unit estimate parameters to calculate the approximate position of the object~\cite{cinefra2014adaptive}. The Received Signal Strength Indicator~(RSSI) is usually collected for distance estimation. The RSSI values can be part of the metadata transmitted with BLE advertisements, thus supporting the positioning system in a simple manner. Such applications only rely on connectionless BLE communications and therefore consume very limited energy. Typically, a set of BLE Beacons deployed in a workspace can support positioning services for several years without the need to change battery.

\begin{figure}[htbp]
\centering
\includegraphics[width=2.5in]{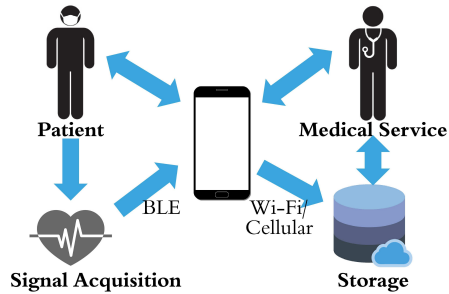}
\caption{Wireless health monitoring system structure~\cite{omre2010bluetooth}.}
\label{fig_health_care}
\end{figure}

The advent of BLE has also attracted considerable interest in the development of personal or human-centric networks. The continuous transmission of body sensor readings usually requires low latency and low power. Figure~\ref{fig_health_care} shows the general structure of a wireless health monitoring system, where BLE can easily connect a set of body sensors to the smartphone by forming a piconet. For easier development, the GATT profile provides a list of identifiers specifically for health service data, such as \textit{blood pressure}, \textit{glucose}, \textit{heart rate}, and \textit{pulse oximeter}. For the health monitoring applications that are primarily interested in body area ($< 1.5 m$) communications, the inherent BLE features satisfy most of the technical requirements with regard to latency, energy consumption, and privacy protection.

As a sub-domain of IoT, home automation is one of the major drivers for the design of BLE. Applications in this domain are usually deployed in a limited area ($<15m$) with more frequent interactions between sensors, actuators, and gateways. BLE also defines a set of GATT services related to home monitoring data, providing simple and direct data extraction for common data types. By virtue of its low-power consumption, BLE has been implemented by many sensors and actuators~\cite{khan2016context}, replacing other wireless technologies, such as Wi-Fi or ZigBee. Some architectures~\cite{collotta2015bluetooth, collotta2015novel, galinina2015smart} also apply BLE as a key part of the gateway in smart homes, where BLE will play a central role in coordinating operation of sensors and actuators.
 
\subsection{Emerging Application Domains}
Apart from the above applications that can be directly supported by BLE, there are numerous other applications that can benefit from BLE-based communications. Most applications in these domains are currently supported by other wireless technologies, but BLE can provide complementary features to, or expand functionality of systems using hybrid networking technologies.

\subsubsection{Smart Infrastructure}

Smart infrastructure is the cornerstone for future smart city development. It intelligently connects energy systems, buildings, and industries to adapt and evolve the way we live and work. We have reached a state of ``infrastructure maturity", especially in developed economies, where the value of new infrastructure is far outweighed by the value of existing infrastructure~\cite{smart_infra}. Applying digital technologies to existing infrastructures offers the potential to use our assets more intelligently, and better meeting social needs. Wi-Fi and cellular solutions have provided stable and robust services in this field, but when it comes to local services, BLE can make it much easier for people to actively interact with buildings and other infrastructure in their surroundings. Such BLE-based services are more responsive, even without Internet connectivity. Examples include: 1) identifying certain groups of people in public areas~\cite{kim2016navigating}, 2) logging and managing the usage and service status of power grids~\cite{kulkarni2019implementing}, and 3) monitoring extreme weather conditions for farming automation~\cite{rajagopal2014low}.

Take smart buildings as an example; here, a proof-of-concept work called NomaBlue~\cite{boukhechba2017novel} uses buildings as databases for spatial exploration in smart city scenarios. This system utilizes both beacons and regular BLE transmitters for data dissemination and device detection. Buildings equipped with BLE beacons and servers act as the knowledge base that stores spatial knowledge from users. Users with BLE transmitters can share spatial knowledge with other users or buildings when in their proximity. Spatial knowledge is then shared and propagated via numerous meet-ups, allowing the system to provide on-demand knowledge requests for local users. The system has been tested and evaluated without Internet connection, and has been proven effective. 
Google's Sidewalk Lab has also implemented similar ideas in the city of Toronto~\cite{sidewalk_labs}. The city-wide deployment of BLE transmitters allow citizens to be aware of people and events in the neighborhood, thus providing opportunities for more social connections.

Using BLE in smart infrastructures can complement other wireless technologies by providing easy and low-cost data transmission in localized areas, even without Internet connection. In case of emergency, for example, BLE may offer an alternative low energy consumption solution monitor peoples location inside buildings and guide evacuations, as everyone will be trying to save battery on their devices and will theoretically have no communication. We therefore see the potential of BLE to be used widely in future infrastructures to build a dynamic and low-cost connected ecosystem. 


\subsubsection{Vehicular Ad Hoc Networks (VANETs)}

VANETs are considered as one of the most prominent technologies for improving the efficiency and safety of modern transportation systems. In a VANET, vehicle-to-vehicle~(V2V) and vehicle-to-infrastructure~(V2I) communication have attracted a lot of interest in the research community. In 2014, Frank et al.~\cite{frank2014bluetooth} first proposed to use BLE as an alternative technology for V2V communications. The idea was to utilize connection-based BLE communications, where each vehicle is initially assigned a role of either central (C) or peripheral (P) and exchanges data with neighbors via C-P connections. For the same role communication~(i.e., C-C or P-P), one of the vehicle will alter its role to set up the connection. This idea has been implemented with a prototype~\cite{bronzi2016bluetooth} using BLE-equipped phones, and tested for single-hop and multi-hop communications.

Yang et al.~\cite{yang2017using, yang2017bluenet} further proved and validated a feasible approach (called BlueNet) for a dynamic and self-organized vehicular network. Their method encapsulated metadata into BLE advertisements so that each node in the network can decide whether to initiate a connection based on the received metadata. BlueNet also applied a dynamic role switching mechanism to provide each node with a higher chance of being connected into the network. Their work also discusses the potential of transferring BlueNet to V2I communications.

The wide deployment of Bluetooth chips in vehicles would greatly lower the cost of large-scale deployment of V2V and V2I systems based on BLE, when compared to technologies such as Dedicated Wireless Short-Range Communication~(DSRC). Therefore, BLE will be more prominent in resource-constraint situations, such as Vehicle-to-Pedestrian~(V2P) and Vehicle-to-Drones~(V2D) communications. Some preliminary models~\cite{wu2017ble, park2017performance} have been developed recently for pedestrian protection using connectionless BLE communication. In these models, the GPS, speed, and direction of pedestrian and vehicles will be broadcast via BLE advertisement, and algorithms are developed based on the received data to predict collision risk. Based on these promising efforts and results in VANETs, BLE can now also be a candidate for communication between other devices such as drones~\cite{komarov2016system, lodeiro2017secure} and robots~\cite{scheunemann2017bluetooth}, e.g., in the context of Mobile Ad Hoc Network (MANET) applications.

BLE may not be able to outperform other wireless technologies due to its limited communications range and lack of diversity in supported types of network topology, but it can serve as a decent complementary technology in a system that utilizes a hybrid networking approach. Table~\ref{vanet_comp} shows a comparison between several existing wireless technologies in VANETs. Though facing challenges in communication range and data rate, the newly released BLE mesh specification~\cite{bluetooth2017mesh} still makes BLE competitive with respect to low cost and easy deployment when compared to Wi-Fi and DSRC.

\begin{table}[htbp]
\centering
\caption{A Comparison of VANET Technologies~\cite{frank2014bluetooth}.}
\label{vanet_comp}
\begin{tabular}{llll}
\hline
 & \textbf{BLE} & \textbf{Wi-Fi} & \textbf{DSRC} \\ \hline
Bandwidth (MHz) & 2 & 20/40 & 10 \\ \hline
Num. of Ch. & 40 & 11 & 7 \\ \hline
Data rate (Mb/s) & 1 & 600 & 27 \\ \hline
Max Power (mW) & 10 & 100 & 2000 \\ \hline
Range (m) & 50 & 100 & 1000 \\ \hline
Latency (ms) & 6 & 50 & 1 \\ \hline
\end{tabular}
\end{table}

\subsubsection{Mobile Payments}
Current mainstream mobile payment methods (i.e., Near Field Communications~(NFC), chip cards, carrier billing, etc.) are typically limited to contact-based or very short range contact-less authentication, while BLE can provide more flexible payment experiences for consumers. For instance, using BLE beacons it is possible to determine the customer's location in a store. From the seller's perspective, it is useful to be aware of a customer's activity - entering the store or checking out. Based on current BLE functions, three different payment scenarios can be presented, of which the second scenario is based on an actual implementation by PayPal:

\begin{compactitem}
    \item Replacing NFC with BLE - pay at cashier with agreement on handset.
    \item PayPal Beacon Hands Free~\cite{todasco2015systems} - pay at cashier with verbal confirmation.
    \item ``Take and Shake'' - scan and pay on your own, no cashier needed.
\end{compactitem}

Unlike the simple scheme of establishing a BLE connection, BLE-based mobile payment requires a more secure process of authentication, which applies to most BLE payment scenarios. Take the approach in~\cite{baldie2016system} as an example, where the payment process usually consists of the following steps: 1) the merchant terminal receives BLE identifiers from the product and the customer, then transmits merchant information related to the product back to the customer device; 2) the customer device sends a request for making a payment; 3) the merchant terminal initiates a transaction based on the product and customer information; 4) the customer device receives and sends back a confirmation to authorize the transaction and the amount; 5) the merchant terminal replies with the confirmation once the transaction is completed. Certain systems may also include two factor authentication approaches to secure the process. The recently opened Amazon Go store is believed to incorporate BLE as part of the solution for automatic checkout~\cite{klinglmayr2017sustainable}. 

As BLE has already been applied in VANET applications, it may also be adapted for scenarios where vehicles can be utilized for mobile payments (e.g., parking fees, fueling, tolls, etc.). The VANET application layer exchanges the vehicle's information with pay station, and then the quick transaction can be processed by a secure BLE transaction layer. While some attempts of using DSRC in such systems have been shown successful~\cite{tinskey2017wireless}, it would be possible to build such solutions also with BLE.

\subsubsection{Multimedia Streaming}
There is also an increasing interest in BLE in applications that require larger data exchange. The data payload limit of BLE in connectionless mode is 31 bytes in version 4.1 and 255 bytes in version 4.2 and after. Although the connection-based mode does not have such a limit on data packets, large amounts of data may still need to fit into the GATT service and characteristics format. For certain applications, such as multimedia streaming, the ability to efficiently stream large amounts of data is an important criterion.

Giovanelli et al.~\cite{giovanelli2015bluetooth} proposed a data streaming design that adds a service composed of two specific characteristics, one for regular data exchanges and the other is used to control the data stream from the central side, enabling it when set to 1 and disabling it when set to 0. In this case, the central device will only need to check the payload from second characteristic to decide whether or not to receive the next regular incoming data packet, thus narrowing the gap between each packet transmission. 

The size limit of data frames at the Link Layer was also tested and explored in~\cite{meli2015streaming}, where it was shown that with proprietary modifications at upper layers, BLE throughput can go as high as 300 kb/s, which is sufficient for most speech and music transmissions. In another study~\cite{shao2016years}, the potential for BLE beacons to transmit complex data, e.g., images, was demonstrated. The proposed method segments an image into multiple smaller units for beaconing, while utilizing a retrieval algorithm on the receiver end to rebuild the images.

In most data streaming use cases, BLE is inferior to other wireless technologies, such as Classic Bluetooth, Wi-Fi and cellular radios, primarily in terms of achievable data rate. However, it still shows great potential as a back-up method, especially for resource-constraint scenarios, such as emergency communication in a natural disaster.

\section{Solutions and Challenges}\label{sol}
The examples discussed so far strongly benefit from various BLE features, but also present challenges and additional requirements that need to be addressed for the successful adoption of BLE in emerging application domains. Based on our review of various examples from emerging application domains, we further categorize four major and urgent challenges or additional requirements for BLE: 1) the need of {\em enhanced data transmission}, 2) {\em mesh metworking} support, 3) {\em inter-operability} with other wireless technologies, and 4) more reliable {\em privacy and security} protection.

\begin{figure}[htbp]
    \centering
    \includegraphics[width=\columnwidth]{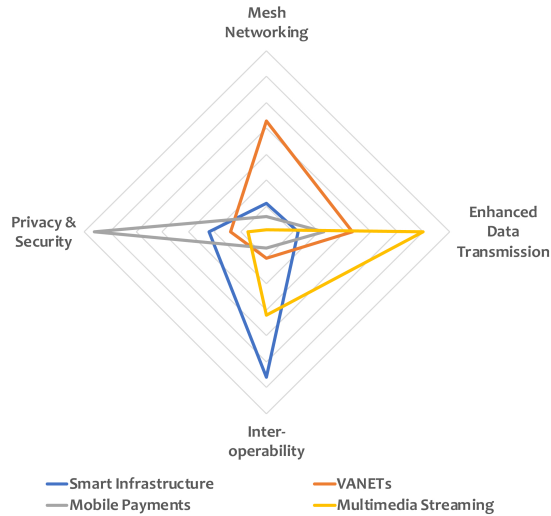}
    \caption{Critical Challenges for Emerging BLE Application Domains.}
    \label{fig:app2sol}
\end{figure}

The spider web map in Figure~\ref{fig:app2sol} demonstrates the significance of these challenges for each of the emerging application domains we reviewed in Section~\ref{app}. 
Smart infrastructure applications usually require interactions and cooperation with other wireless devices, and also need some protection for the transmission of sensitive data. 
VANET applications, in contrast, exhibit a strong need for mesh networking capabilities, e.g., to connect as many vehicles as possible and to exchange information between these vehicles in a timely fashion. An enhanced data transmission mechanism is expected to further accommodate the rapidly growing vehicular networks, while also improving their throughput.
For mobile payments, a main challenge for BLE is privacy protection, where the simplicity of the BLE design can lead to a vulnerability to attacks. Multimedia streaming applications require higher data rates and lower latencues for BLE, while also having a need for inter-operability for use cases where BLE works as an auxiliary approach for other technologies.

While some of these challenges have been addressed by prior work, not all have yet received attention or have been fully resolved, indicating the necessity of further research into these areas. In this section, we summarize and analyze some solution approaches to the challenges mentioned above.

\subsection{Enhanced Data Transmission}
The two types of BLE communications between two devices (i.e., connection-based and connectionless) have enabled simple and quick data exchange for many novel designs and applications. However, there are still some challenges found in application domains such as IPS and VANETs, where BLE encounters several limitations in device discovery and role assignment. We list and analyze several typical solutions below.

\subsubsection{Efficient Device Discovery}
In most prior studies, it is assumed that the advertisements are processed immediately as long as they are received by a scanner, which means there is no collision when discovering BLE devices. However, in practice, there exist contentions among multiple BLE devices during the device discovery process, especially in a crowded environment. With an increasing number of BLE devices, discovery latency and energy consumption will rise exponentially~\cite{cho2016performance}, which will degrade the achievable performance and QoS of many applications. Solutions for low-latency device discovery fall into two categories: 1) {\em Adaptive Parameter Settings} and 2) {\em Channel Sensing}, which are further explained below.

{\em Adaptive Parameter Settings}: There are four major timing parameters that may affect the performance of BLE device discovery:
\begin{compactitem}
    \item $AdvInterval$: total time of an advertising event on three advertising channels;
    \item $AdvTimePerChannel$: advertising period per channel;
    \item $ScanWindow$: time length of a scan event;
    \item $ScanInterval$: delay between two consecutive scans.
\end{compactitem}
Although the BLE standard has enabled the ability for BLE devices to operate with a wide range of these parameters, it is somehow hard for them to fine tune these parameters automatically, as neither the scanner nor the advertiser can be aware of the existence of contentions on advertising channels. The adaptive parameter setting method adjusts some of these parameters based on certain observations to achieve a more efficient device discovery. The authors in \cite{liu2013adaptive} presented a model that utilizes the advertisement payload to reflect contention situations. The main idea can be summarized as: 1) advertisers piggyback connection reports (containing $AdvInterval$ and the current count of advertising intervals) in the advertisement payload; 2) scanners use the reports to evaluate the current contention situation and adjust certain parameters ($ScanWindow$ and $ScanInterval$) accordingly. However in practical use, especially when the parameters are changing fast, the advertiser needs to frequently update the payload, which may lead to high system overhead.

Park et al.~\cite{park2016adaptive} provided a parameter adjustment scheme based on the discovery time ratio $\rho$ ($= T_{actual} / T_{reference}$). The reference discovery time is defined according to empirical settings. Advertisers and scanners will then adjust $AdvInterval$ and $ScanWindow$, respectively, depending on $\rho$. This approach allows advertisers to actively join the parameter adjustment, which makes it easier to be applied to general use cases.

{\em Channel Sensing}: The latency in the BLE device discovery is partly caused by the unawareness of channel traffic. Therefore, the main idea of the Channel Sensing method is to sense the channel traffic before advertising or scanning. When the channel is busy, an advertiser will execute a deferral and random backoff and retry the sensing until the channel is idle~\cite{seo2016discovery}. A similar scheme can also be applied to scanners so that they can decide whether to initiate a connection to neighbors based on channel traffic~\cite{kim2017backoff}. The channel traffic can be evaluated by the density of the received packets, but may experience heavy fluctuations in highly dynamic networks. While promising, this method still requires a better definition of the channel traffic and the decision-making algorithm for each role.

The two types of methods mentioned above have shown the effectiveness in reducing latency caused by collisions in certain cases. However, the common problem with these solutions is the lack of tests and validation on physical BLE devices. Device discovery is primarily conducted in the physical layer and link layer, which means that the enhancements will need to manipulate the BLE stack and must be useful for various common applications. Further, the schemes described above are mainly designed and tested for dense networks and the trade-offs when applied to sparse networks have not been investigated.

\subsubsection{Role Assignments}
In many applications, independent devices will be expected to find ways to coordinate, access each others’ sensor data, share communication channels, or process and fuse data from multiple sources. Self-organization and self-management of these devices, e.g., the ability to decide when to establish or tear down a connection, is one of the defining characteristics of IoT and VANETs. However, BLE was built for deployment scenarios that have well-defined roles for the connected devices, e.g., it is defined that only a central device can initiate a connection to peripherals. This design may limit the self-management of BLE device and the efficiency of data propagation. Existing solutions to this limitation focus on providing alternatives for the role assignment in BLE connections. 

Bronzi et al.~\cite{bronzi2016bluetooth} proposed an event-based role switching scheme to switch the role of central and peripheral that a BLE device can play, so that each device can be in the accessible role when there is a need to transfer data. The scheme is designed in the context of multi-hop V2V communications, where every node (vehicle) is initially set as a central and will activate its peripheral role manually on the device itself or upon receiving a message to rebroadcast. In a multi-hop scenario, the relay node will stay active as both central and peripheral roles for a certain period of time to ensure the data integrity, and will resume the initial role once the data transmission is finished. However, the transmission latency may increase aggressively as the network size and the amount of data grows.

Following a similar idea of role switching, Yang et al.~\cite{yang2017using, yang2017bluenet} further developed a more self-organized mode of role switching, which is called {\em Dynamic C/P Switching}. This approach addresses the problem by allowing BLE devices to continually and frequently switch their roles. In general, if the roles are switched at random times, two devices will eventually have different roles, allowing them to establish a connection and communicate. In this mode, a device may be initially be set as either central or peripheral and after a random period of time, its role will be switched to the other. The {\em active time} for each role is randomly chosen, possibly from a range of minimum and maximum active times. At any given time, devices with the same role are not able to make a connection and a connection can only be established if one device is central, while the other is peripheral. Once a connection has been established, the role switching process can be paused to allow the newly connected devices to exchange their data. When the data transfer has finished, the connection will be torn down and the switching approach will resume. This method allows equal opportunities for each device to play as central or peripheral, but the frequent role switching may also bring high overhead and low energy-efficiency.  

Similar ideas were also applied to the data flow between parent nodes and child nodes in a BLE-based tree networks~\cite{kim2015bluetooth}. However, these role switching schemes face some common challenges: 1) how to provide an efficient parameter settings (e.g., the time interval for each role) according to different network contexts; 2) how to minimize the overhead caused by role switching, particularly in dense networks; and 3) how to ensure the consistency of data transmissions.

\subsection{Mesh Networking}
The limited range of BLE is one of its major drawbacks, especially for use cases in health monitoring, IoT, and VANETs. BLE was originally designed based on the star network topology. However,  other wireless technologies, such as ZigBee and Z-Wave, support mesh networks, which provides them with an advantage in these domains~\cite{darroudi2017bluetooth}. Therefore, to further extend the network coverage of BLE, solutions have been proposed, both by the community and the Bluetooth SIG.

In mid-2017, the Bluetooth SIG released a set of specifications defining the architecture of the BLE mesh topology~\cite{bluetooth2017mesh}. These specifications outline the implementations and requirements to enable an interoperable many-to-many mesh networking solution on top of Bluetooth 5.0 and later. Apart from the official support for mesh networking, many solutions from academia have also been proposed based on Bluetooth 4.0 through 4.2 to fit into multiple application domains. Based on the type of routing protocol, these solutions can be categorized into three classes: 1) {\em Flooding routing}, 2) {\em Table-driven routing} and 3) {\em On-demand routing}. While all of them resolve the routing problems in BLE mesh networks, the target application domains may vary. Smart infrastructure applications are usually deployed in a static environment, where flooding and table-driven routing may exhibit high performance. Similarly, built on a pre-defined and well connected network, health monitoring applications can easily benefit from table-driven routing methods. However in VANETs use cases, the network is usually of high mobility and flexibility, where flooding and table-driven routing are less efficient but on-demand routing can provide more dynamic routes.

We review and analyze typical solutions from each category. The main objectives and approaches for each of the solutions are explained below.

\subsubsection{Flooding Routing}
This type of routing protocols are based on connectionless broadcast, where every node that receives a broadcast will rebroadcast until the message has been received by the destination node. {\em BLESSED}~\cite{turkes2015blessed} is an example that utilizes the BLE advertisement for flooding packets. It defines three superstates for each participating node, so that every node will be able to broadcast, receive, and update identifiers of advertisements. However, the BLESSED method needs to be implemented and tested with the assistance of Wi-Fi hotspots.

{\em BLEmesh}~\cite{kim2015blemesh}, proposed in the same year, is a pure BLE-based mesh solution. In BLEmesh, packets carrying data from a specific source-destination couple are aggregated in batches. Data, together with control fields, which are used to decide which nodes will participate as broadcasters, are carried in the advertisements. The control fields include two lists: 1) {\em Forwarder List} and 2) {\em Batch Map}, which are used to keep track of a prioritized set of intermediate nodes and the last nodes that have broadcast data to a corresponding batch, respectively. These control fields are designed to minimize the overhead caused by unnecessary rebroadcast. The authors compared their protocol with a conventional flooding protocol and show that BLEmesh requires fewer transmissions.

\subsubsection{Table-driven Routing}
Table-driven routing is a typical type of routing in MANETs. In such protocols, each node maintains one or more tables containing routing information to every other node in the network. {\em MHTS}~\cite{mikhaylov2013multihop} is known as the first attempt to construct a multi-hop BLE network. This approach is based on BLE GATT profiles, which stores the routing entries (including data, source address, and destination addresses). The route discovery is carried out by broadcasting the {\em Seek Table} in advertisements and constructing a {\em Route Table} for each receiving node. Then connections will be established based on the Route Table. Since Bluetooth 4.0 does not allow scatternet formations, current connections must be torn down before the intermediate node transfers data to the next hop. MHTS can transfer packets over up to five hops for a file size of 1 kB.

{\em BMN}~\cite{sirur2015mesh} was later proposed as an improved table-driven routing protocol. BMN uses Directed Acyclic Graph~(DAG) as the basis for routing. The network formation starts with BLE advertisements of DAG information and an arbitrarily selected root node. Then a DAG will be formed by a series of connection establishments, and each connected node will maintain a table that stores its parent, alternative parent, and its children. The data forwarding will then retrieve its referenced path from the table maintained by each node. However, this approach may suffer from intermediate node failure and network congestion.

{\em NDN-BLE}~\cite{balogh2015service} is a variation of table-driven routing protocol, which uses the Named Data Network~(NDN) to build BLE mesh networks. NDN names every chunk of data with an appropriate Uniform Resource Identifier (URI), and operates over a distributed database, which determines how an endpoint can retrieve data of interest. A Mediation Service is introduced to manage and aggregate distributed databases, so that multi-hop transmissions are actually conducted via these mediators. While this approach has been verified from various angles, the authors did not provide any physical or simulation tests to evaluate its performance. 

\subsubsection{On-demand Routing}
On-demand routing is another type of routing protocols in MANETs. In contrast to table-driven routing protocols, routes are not maintained at every node, instead the routes are created as and when required. When a source wants to send data to a destination, it invokes the route discovery mechanisms to find a path to the destination. The route remains valid until the destination is not reachable anymore or until the route is no longer needed. Guo et al.~\cite{guo2015demand} present a typical example of this type of routing based on BLE. In their method, the route discovery is conducted based on scatternets: a source node first sends a route request to its master. If the destination node is not in the slaves list, then the master starts a Breadth-First Search by forwarding the request to its slaves that are also in other piconets. Such slaves will continue the same procedure until the destination node is found. The source node will exploit the shortest path once all possible routes are obtained. This approach may waste network resources as only one route is used and no other nodes benefit from it.

Therefore, {\em CORP}~\cite{jung2017topology} was proposed to provide a more effective routing procedure. The scatternet formation in CORP is based on a device's degree (i.e., $\# neighbors$) so that the each piconet will be connected with the smallest number of connections. The routing protocol is executed similar to the previous method, but due to the simplification of scatternets, the route discovery is now more efficient. The authors also design a mechanism to locally reconstruct the scatternets when encountering a single-node failure.

{\em RT-BLE}~\cite{patti2016bluetooth} and {\em MRT-BLE}~\cite{leonardi2018multi} stabilize the routing by putting limitations on each node: 1) a node can establish a connection with up to two masters and 2) a master can establish a connection with at most one other master, and, in this connection, the former shall play the slave role. In the route discovery process, these two models use the Client Characteristic Configuration Descriptor (CCCD), a descriptor available in the GATT layer, to maintain connections. CCCD only allows one connection to be active at a time, while the rest are kept inactive. Such models show improvements in power consumption, while achieving the goal of multi-hop transmissions. Other variations (e.g., Dual-Ring Tree~\cite{yu2018reliable}) follow similar approaches to on-demand routing, but use tree topologies to construct the network, which have also been proven useful in some cases.

\subsubsection{Comparison}
Table~\ref{mesh_comp} summarizes the typical solutions for each of the afore mentioned routing classes and also lists several key characteristics for each solution. Since each work uses different metrics and different hardware/chipsets for evaluating the performance, the comparison may vary depending on the actual test environment. Generally, the Latency in Table~\ref{mesh_comp} refers to the general latency for multi-hop transmissions, while the Dissemination Efficiency reflects the average packet delivery ratio in a mesh network. These key characteristics are generally acknowledged factors that need to be considered when designing a network-based application. There is no solution that fits into all design criteria, but we can see from the table that the BLESSED in flooding routing, BMN in table-driven routing and CORP in on-demand routing stand out for delivering on most of the criteria. We suggest developers should refer to the best fit based on their actual goal/needs. 

\begin{table*}[htbp]
\centering
\caption{A comparison of BLE mesh network solutions.}
\label{mesh_comp}
\resizebox{\textwidth}{!}{%
\begin{tabular}{cccccccccc}
\hline
\multicolumn{1}{c|}{\multirow{2}{*}{\textbf{\begin{tabular}[c]{@{}c@{}}Routing\\ Type\end{tabular}}}} & \multicolumn{1}{c|}{\multirow{2}{*}{\textbf{\begin{tabular}[c]{@{}c@{}}MajorApplication\\ Domains\end{tabular}}}} & \multicolumn{1}{c|}{\multirow{2}{*}{\textbf{\begin{tabular}[c]{@{}c@{}}Solution\\ Name\end{tabular}}}} & \multicolumn{1}{c|}{\multirow{2}{*}{\textbf{Year}}} & \multicolumn{1}{c|}{\multirow{2}{*}{\textbf{\begin{tabular}[c]{@{}c@{}}Bluetooth\\ Version\end{tabular}}}} & \multicolumn{5}{c}{\textbf{Key Characteristics}} \\ \cline{6-10} 
\multicolumn{1}{c|}{} & \multicolumn{1}{c|}{} & \multicolumn{1}{c|}{} & \multicolumn{1}{c|}{} & \multicolumn{1}{c|}{} & \multicolumn{1}{c|}{\textbf{Latency}} & \multicolumn{1}{c|}{\textbf{\begin{tabular}[c]{@{}c@{}}Energy \\ Efficiency\end{tabular}}} & \multicolumn{1}{c|}{\textbf{\begin{tabular}[c]{@{}c@{}}Dissemination \\ Efficiency\end{tabular}}} & \multicolumn{1}{c|}{\textbf{Mobility}} & \textbf{Scalability} \\ \hline
\multicolumn{1}{c|}{\multirow{2}{*}{\textit{Flooding}}} & \multicolumn{1}{c|}{\multirow{2}{*}{Smart Infrastructure,}} & \multicolumn{1}{c|}{BLEmesh~\cite{kim2015blemesh}} & \multicolumn{1}{c|}{2015} & \multicolumn{1}{c|}{4.2} & \multicolumn{1}{c|}{} & \multicolumn{1}{c|}{\Checkmark} & \multicolumn{1}{c|}{\Checkmark} & \multicolumn{1}{c|}{} &  \\ \cline{3-10} 
\multicolumn{1}{c|}{} & \multicolumn{1}{c|}{} & \multicolumn{1}{c|}{BLESSED~\cite{turkes2015blessed}} & \multicolumn{1}{c|}{2015} & \multicolumn{1}{c|}{4.2} & \multicolumn{1}{c|}{} & \multicolumn{1}{c|}{\Checkmark} & \multicolumn{1}{c|}{\Checkmark} & \multicolumn{1}{c|}{\Checkmark} & \Checkmark \\ \hline
\multicolumn{1}{c|}{\multirow{3}{*}{\textit{Table-driven}}} & \multicolumn{1}{c|}{\multirow{3}{*}{\begin{tabular}[c]{@{}c@{}}Smart Infrastructure,\\ Health Monitoring\end{tabular}}} & \multicolumn{1}{c|}{MHTS~\cite{mikhaylov2013multihop}} & \multicolumn{1}{c|}{2013} & \multicolumn{1}{c|}{4.0} & \multicolumn{1}{c|}{\Checkmark} & \multicolumn{1}{c|}{} & \multicolumn{1}{c|}{} & \multicolumn{1}{c|}{} &  \\ \cline{3-10} 
\multicolumn{1}{c|}{} & \multicolumn{1}{c|}{} & \multicolumn{1}{c|}{NDN-BLE~\cite{balogh2015service}} & \multicolumn{1}{c|}{2015} & \multicolumn{1}{c|}{4.1} & \multicolumn{1}{c|}{} & \multicolumn{1}{c|}{} & \multicolumn{1}{c|}{\Checkmark} & \multicolumn{1}{c|}{} &  \\ \cline{3-10} 
\multicolumn{1}{c|}{} & \multicolumn{1}{c|}{} & \multicolumn{1}{c|}{BMN~\cite{sirur2015mesh}} & \multicolumn{1}{c|}{2015} & \multicolumn{1}{c|}{4.1} & \multicolumn{1}{c|}{\Checkmark} & \multicolumn{1}{c|}{\Checkmark} & \multicolumn{1}{c|}{\Checkmark} & \multicolumn{1}{c|}{\ding{55}} &  \\ \hline
\multicolumn{1}{c|}{\multirow{5}{*}{\textit{On-demand}}} & \multicolumn{1}{c|}{\multirow{5}{*}{VANETs}} & \multicolumn{1}{c|}{N/A~\cite{guo2015demand}} & \multicolumn{1}{c|}{2015} & \multicolumn{1}{c|}{4.1} & \multicolumn{1}{c|}{\Checkmark} & \multicolumn{1}{c|}{} & \multicolumn{1}{c|}{} & \multicolumn{1}{c|}{\ding{55}} &  \\ \cline{3-10} 
\multicolumn{1}{c|}{} & \multicolumn{1}{c|}{} & \multicolumn{1}{c|}{CORP~\cite{jung2017topology}} & \multicolumn{1}{c|}{2017} & \multicolumn{1}{c|}{4.1} & \multicolumn{1}{c|}{\Checkmark} & \multicolumn{1}{c|}{\Checkmark} & \multicolumn{1}{c|}{\Checkmark} & \multicolumn{1}{c|}{\Checkmark} & \Checkmark \\ \cline{3-10} 
\multicolumn{1}{c|}{} & \multicolumn{1}{c|}{} & \multicolumn{1}{c|}{RT-BLE~\cite{patti2016bluetooth}} & \multicolumn{1}{c|}{2016} & \multicolumn{1}{c|}{4.1} & \multicolumn{1}{c|}{\Checkmark} & \multicolumn{1}{c|}{} & \multicolumn{1}{c|}{} & \multicolumn{1}{c|}{\ding{55}} & \ding{55} \\ \cline{3-10} 
\multicolumn{1}{c|}{} & \multicolumn{1}{c|}{} & \multicolumn{1}{c|}{Dual-Ring Tree~\cite{yu2018reliable}} & \multicolumn{1}{c|}{2018} & \multicolumn{1}{c|}{N/A} & \multicolumn{1}{c|}{} & \multicolumn{1}{c|}{} & \multicolumn{1}{c|}{\Checkmark} & \multicolumn{1}{c|}{\Checkmark} & \Checkmark \\ \cline{3-10} 
\multicolumn{1}{c|}{} & \multicolumn{1}{c|}{} & \multicolumn{1}{c|}{MRT-BLE~\cite{leonardi2018multi}} & \multicolumn{1}{c|}{2018} & \multicolumn{1}{c|}{4.1} & \multicolumn{1}{c|}{\Checkmark} & \multicolumn{1}{c|}{} & \multicolumn{1}{c|}{\Checkmark} & \multicolumn{1}{c|}{} &  \\ \hline
\multicolumn{10}{l}{\Checkmark: theoretically or experimentally proved to be good; \ding{55}: weakness or proved to be below average; blank: not considered or explained explicitly.}
\end{tabular}%
}
\end{table*}

In summary, BLE mesh networking is an emerging area with the potential to expand the BLE applicability space. The above solutions can address various application scenarios, but many of them still suffer from trade-offs between latency, network traffic, and power consumption when applied to general cases and, therefore, this area requires further exploration.

\subsection{Inter-operability}
Although cheap and convenient, BLE cannot yet fully replace other existing wireless technologies. In practical uses, it will be more common to see BLE co-exist or even integrate with technologies that share the 2.4 GHz frequency band. This brings up the problem of inter-operability, i.e., approaches to co-existence and integration with other technologies that minimize inter-technology interference. Next, we discuss such challenges in terms of co-existence and integration.

\subsubsection{Co-existence}
BLE shares a similar frequency band with Wi-Fi, ZigBee, and many other wireless technologies. Silva et al.~\cite{silva2014coexistence} have performed comprehensive tests on co-existence and interference among BLE, Classic Bluetooth, Wi-Fi, and ZigBee in a full anechoic chamber. RSSI and Bit Error Rate are measured as the main criteria for the level of interference. The results show no obvious interference between BLE and Wi-Fi, but Classic Bluetooth and ZigBee do collide occasionally with BLE, as shown by the increased RSSI and Bit Error Rate. Kalaa et al.~\cite{al2016evaluating} presented a general evaluation of BLE in realistic wireless environments such as a sports facility, a university's food court, and a hospital's Intensive Care Unit~(ICU). Their results revealed that the probability of failed BLE connections and transmissions increases sharply when the wireless environment become extremely complex and dense. The PHY layer of BLE stack utilizes an adaptive channel hopping scheme to avoid interference, and it has been shown effective in most cases. However, the evaluation results above also point out that there is a need for additional research to develop enhanced schemes for more critical and complex scenarios. 

\subsubsection{Integration}
We expect that, due to its limitations with respect to data rate and communication range, BLE will often be used together with other wireless technologies to achieve certain goals and functionality.

A common method of integration is to utilize the simple and fast connection process of BLE to set up connection between devices, while the actual data transmissions are performed via another wireless technology. For example, this approach has been implemented and tested for Wi-Fi P2P communications~\cite{joh2015hybrid}, where a pair of nodes exchange Wi-Fi MAC addresses and service lists over BLE advertisements. Nodes then decide whether to establish a Wi-Fi P2P connection based on the contents of the received BLE messages. Once a connection has been established, all data exchange occurs via the Wi-Fi P2P channels. This approach can significantly shorten the latency of connection establishment, while maintaining a high transmission rate. A similar approach could also be based on Classic Bluetooth.

Another type of integration is to transmit non-BLE data packets over BLE networks. One popular idea is to transmit Internet Protocol version 6~(IPv6) packets over BLE. The Internet Engineering Task Force~(IETF) published the ``IPv6 over Bluetooth Low Energy'' specification in 2015~\cite{nieminen2015ipv6}, formally defining a complete protocol stack that integrates BLE~(Bluetooth version 4.1 or greater) and IPv6. This stack is based on the {\em IPv6 over Low-Power Wireless Personal Area Network~(6LoWPAN)} standard, which is also maintained by IETF and intended for transmitting IPv6 packets over low-power networks. The IPv6 stack (including 6LoWPAN, IPv6, and UDP/TCP) and GATT stack~(including GATT and ATT) work in parallel on top of the BLE L2CAP layer. The 6LoWPAN and L2CAP layers provide address configuration, header compression, fragmentation, and reassembly services to translate IPv6 packets into transferable packets for the BLE PHY layer and vice versa. Based on this concept, several architectures were proposed that yield high throughput~\cite{yim2015ipv6} or seamless transitions between different protocols~\cite{wang2013transmitting}.

Overall, integrating BLE with existing wireless technologies is a method that can be used to overcome certain weaknesses of each technology, while also offering better compatibility among different types of devices. However, there is still a dearth of work on effective and efficient solutions using this concept, including efforts to standardize such solutions.

\subsection{Privacy and Security}
Security is of the utmost importance in many application domains. Although recent revisions of BLE were designed for better security and have fixed several vulnerabilities~\cite{ryan2013bluetooth,das2016uncovering}, security still remains as a significant challenge for many current and future BLE applications. 

Ray et al. successfully performed a series of attacks on Bluetooth 4.2 devices~\cite{ray2018bluetooth} and revealed the weaknesses that have not yet been covered by the LE Secure scheme. They conducted three major attacks on BLE: MITM, Flooding, and Fuzzing attacks. The MITM attack easily breaks a BLE connection and successfully modifies the data packets. The Flooding and Fuzzing attacks do not significantly crash the target BLE device when numerous connection requests and GATT requests are sent to it, but still cause the target device to be ``stuck'' for a certain period of time. Similar attacks have also been conducted to sniff data from BLE devices~(such as keyboards~\cite{willingham2018testing}). 

Due to the insecure channels and the simplicity of connections, there is a lack of countermeasures that can protect BLE devices from MITM attacks. However, some attempts have been made to secure the transmitted packets, e.g., Perrey et al.~\cite{perrey2011wisec} proposed a key exchange approach that complements the ECDH algorithm to provide secure connections. This approach applies Merkle's Puzzle to the key generation process, where all puzzles are generated and encrypted depending on the functionality of the devices~(i.e., fully functioning or reduced functioning). The encrypted puzzles are then broadcast via BLE advertisements for key distribution, so that further connections will be securely established based on the key. Another attempt, called {\em Black BLE}~\cite{chakrabarty2016black}, applied AES-EAX encryption to both meta-data and payloads of PDUs in each transmission. The encryption method is robust, but it also raises the problem of symmetric key management and reduced payload efficiency.

\section{Discussion and Conclusion}\label{con}

BLE is an emerging wireless technology with great potential in many application areas. Ever since its release, it has been successfully applied in many different domains. This paper surveys several BLE applications and describes the challenges and existing solutions regarding data transmissions, mesh networking, inter-operability, and security.

Among traditional application domains, IPS can directly build upon the connectionless BLE transmission, while health monitoring and Home Automation applications rely on connection-based communications and basic BLE network topology. Emerging application domains require more advanced features, such as the inter-operability and mesh network support for smart infrastructure and VANETs, privacy protection for mobile payments and more robust data transmission for multimedia streaming applications.
With recent enhancements proposed by the research community, we see that there is great potential for BLE and its use in more emerging domains, especially with the rise of Industrial IoT as its ubiquitous applications in static sensor networks as well as in mobility and transportation industries. While the investigations of current solutions show that some progress has been made, we still witness many opportunities and unsolved problems. Specifically, we propose that the community start with the following directions:

\subsection{From Traditional to Emerging }
With the release of Bluetooth 5.1, many BLE features have been improved, providing even more opportunities for traditional BLE applications. Bluetooth 5.1 brings two positioning systems to BLE: Angle of Arrival~(AoA) and Angle of Departure~(AoD). Since they are implemented at the hardware level, BLE-based IPS will then provide more convenient and accurate positioning services even with very few BLE beacons. Based on the location information, home automation can also benefit, e.g., smart home devices can now precisely detect nearby users and take actions accordingly.

Thanks to its low-cost, BLE-based sensors have been widely deployed under various settings, and a lot of data have been generated by these sensors. Utilizing the data and applying machine learning approaches, many more applications could emerge, e.g., attempts have been made to learn and predict moving patterns~\cite{pu2018indoor} and distances~\cite{lam2018improved} from the collected data. Similar technologies can also be applied in the Geo-spatial domain, where BLE positioning services and learning models can be used to draw high definition maps for complex road intersections.

\subsection{From Small to Large}
As BLE technology is now being deployed as a key part in smart city implementation, scalability of the solutions is going to be of paramount interest to the industry. Mesh networking support can apparently expand the range of BLE networks, but there are many other issues that are not fully considered, such as the discovery scheme, connection robustness,  support for large payload size of data packets in multi-hop transmission, etc. Since multicast is not supported by connection-based BLE communications~\cite{chang2014bluetooth}, system latency is expected to increase significantly as the network scales. Therefore, methods that address many of these issues under one holistic solution architecture design, are currently considered to be the most promising ones to scale the BLE deployment.

\subsection{BLE in the era of 5G}
It is expected that the arrival of 5G will provide many new opportunities and although 5G communication supports large bandwidth and high data rate communications, BLE can still play a significant role alongside 5G. For example, applications such as bike sharing and station planning services, last-mile delivery services, brain-machine interfaces, etc., can benefit from BLE's ability to provide localized communications, while 5G can handle any required large-scale data transfers for these applications.

With the rise of the Industrial Internet of Things~(IIoT) and other technologies such as edge computing, a seamless integration of multiple wireless technologies will often be essential for many deployment environments. IP-based methods~\cite{nieminen2015ipv6} are the most popular approaches to connecting constrained devices to the Internet, and it is expected that this will continue to grow due to the ability to provide scalability and simple application development. We therefore expect that BLE will also be used alongside other radio technologies, such are LTE ad 5G, to address the needs of IIoT applications.

\section*{Acknowledgment}

The project was supported through a Ford Motor Company University Research Program (URP) grant.

\section*{References}

\bibliography{BLE_survey}

\end{document}